\documentclass[a4paper,11pt]{article}

\usepackage{contribution}

\newcommand{\weblink}[2][]{%
    \ifthenelse{\equal{#1}{}}%
    {\textnormal{\url{#2}}}%
    {\textnormal{\href{#2}{#1}}}%
}

\def\beq{\begin{equation}}
\def\eeq#1{\label{#1}\end{equation}}
\def\eeqn{\end{equation}}

\def\beqa{\begin{eqnarray}}
\def\eeqa#1{\label{#1}\end{eqnarray}}
\def\eeqan{\end{eqnarray}}

\let\bar=\overbar

\def\Dslash{\not{\hbox{\kern-4pt $D$}}}
\def\dslash{\not{\hbox{\kern-2pt $\del$}}}

\def\msb{{\bar{\ssstyle M \kern -1pt S}}}

\newcommand{\contribution}[7][]{%
  \clearpage
  \thispagestyle{plain}
  \ifthenelse{\equal{#1}{}}
  {\hypersetup{pdftitle={#2}}}
  {\hypersetup{pdftitle={#1}}}
  \hypersetup{pdfauthor={{#3} {#4}}}
  {\centering\normalfont\LARGE\bfseries\sffamily #2 \par\nobreak}
  \lhead{}
  \chead{%
    \textit{\footnotesize XIV International Conference on Hadron Spectroscopy
      (\weblink[\textit{hadron2011}]{http://www.hadron2011.de}), 13-17 June 2011, Munich, Germany}%
  }
  \rhead{}
  \bigskip
  \begin{center}
    {#3} {#4}\ifthenelse{\equal{#6}{}}{}{\footnote{\weblink[#6]{mailto:#6}}}
    \ifthenelse{\equal{#7}{}}{}{#7} \\
    \textit{#5}
  \end{center}
  \bigskip
}

\renewcommand{\abstract}[1]{%
  \begin{center}
    \begin{minipage}{0.85\textwidth}
      \begin{footnotesize}
        #1
      \end{footnotesize}
    \end{minipage}
  \end{center}
  \bigskip
}
\newcommand{\br}[1]{\mathcal{B}(#1)}

\begin{document}

%
%
%
%
%
{  


%

\contribution[Study ]  
{Study of $a^{0}_{0}(980)-f_{0}(980)$ mixing at BESIII}  
{Chunyan}{Liu}  
{Institute of High Energy Physics, Chinese Academy of Sciences, China, 100049}  
{}  
{on behalf of the BESIII Collaboration}  
%

\abstract{%
In this talk,we present direct measurements of $a^{0}_{0}(980)-f_{0}(980)$ mixing in the processes $J/\psi\to\phi f_{0}(980)\to\phi a^{0}_{0}(980)\to\phi\eta\pi^{0}$ and $\psi^{\prime}\to\gamma\chi_{c1},\chi_{c1}\to\pi^{0} a^{0}_{0}(980)\to\pi^{0} f_{0}(980)\to\pi^{0}\pi^{+}\pi^{-}$ with $2.25 \times 10^{8}$ $J/\psi$ data and $1.06 \times 10^{8}$ $\psi^{\prime}$ data at BESIII.
}
%

\section{Introduction}
BESIII/BEPCII~\cite{:2009vd} is a major upgrade of the BESII detector and BEPC accelerator. The primary physics purposes are aimed at the study of hadron spectroscopy and $\tau$-charm physics.\\
The nature of the scalar mesons  $a^{0}_{0}(980)$ and $f_{0}(980)$ has been a hot topic in light hadron physics for many years. The mixing between  $a^{0}_{0}(980)$ and $f_{0}(980)$ is expected to shed light on the nature of these two resonances.
In this talk,we present the recent result from the study of $a^{0}_{0}(980)-f_{0}(980)$ mixing at BESIII~\cite{Yadi2011}.

\section{Study of $a^{0}_{0}(980)-f_{0}(980)$ mixing at BESIII}
The leading contribution to the isospin-violating mixing transition amplitudes for  $f_{0}(980)\to a^{0}_{0}(980)$ and $a^{0}_{0}(980)\to f_{0}(980)$ are shown to be dominated by the difference of the unitarity cut which arises from the mass difference between the charged and neutral kaons. As a consequence, a narrow peak of about 8 $MeV/c^{2}$ is predicted between the charged and neutral kaon thresholds. Using the samples of $2.25 \times 10^{8}$ $J/\psi$ and $1.06 \times 10^{8}$ $\psi^{\prime}$ events,we perform direct measurements of $a^{0}_{0}(980)-f_{0}(980)$ mixing in the processes $J/\psi\to\phi\eta\pi^{0}$ and $\psi^{\prime}\to\gamma\chi_{c1},\chi_{c1}\to\pi^{0} a^{0}_{0}(980)\to\pi^{0} f_{0}(980)\to\pi^{0}\pi^{+}\pi^{-}$.\\\\
FIG.(a) shows the fitting results of $\eta\pi^{0}$ mass spectrum recoiling against $\phi$ signal in $J/\psi\to\phi f_{0}(980)\to\phi a^{0}_{0}(980)\to\phi\eta\pi^{0}$.The dotted line shows the mixing signal. The dash-dotted line denotes the $a^{0}_{0}(980)$ from virtual photon or $K^{*}\bar{K}$ loop. The dashed line is a polynomial background constrained to the $\phi$ sideband. The significance of  $f_{0}(980)\to a^{0}_{0}(980)$ mixing signals is 3.3 $\sigma$. The mixing branching ratio is determined to be $\br{J/\psi\to\phi f_{0}(980)\to\phi a^{0}_{0}(980)\to\phi\eta\pi^{0}}$($<5.4\times 10^{-6}$ at the $90\%$ C.L.).\\
\begin{figure}[htbp]
\begin{center}
  \includegraphics[width=4.5cm,height=4.5cm]{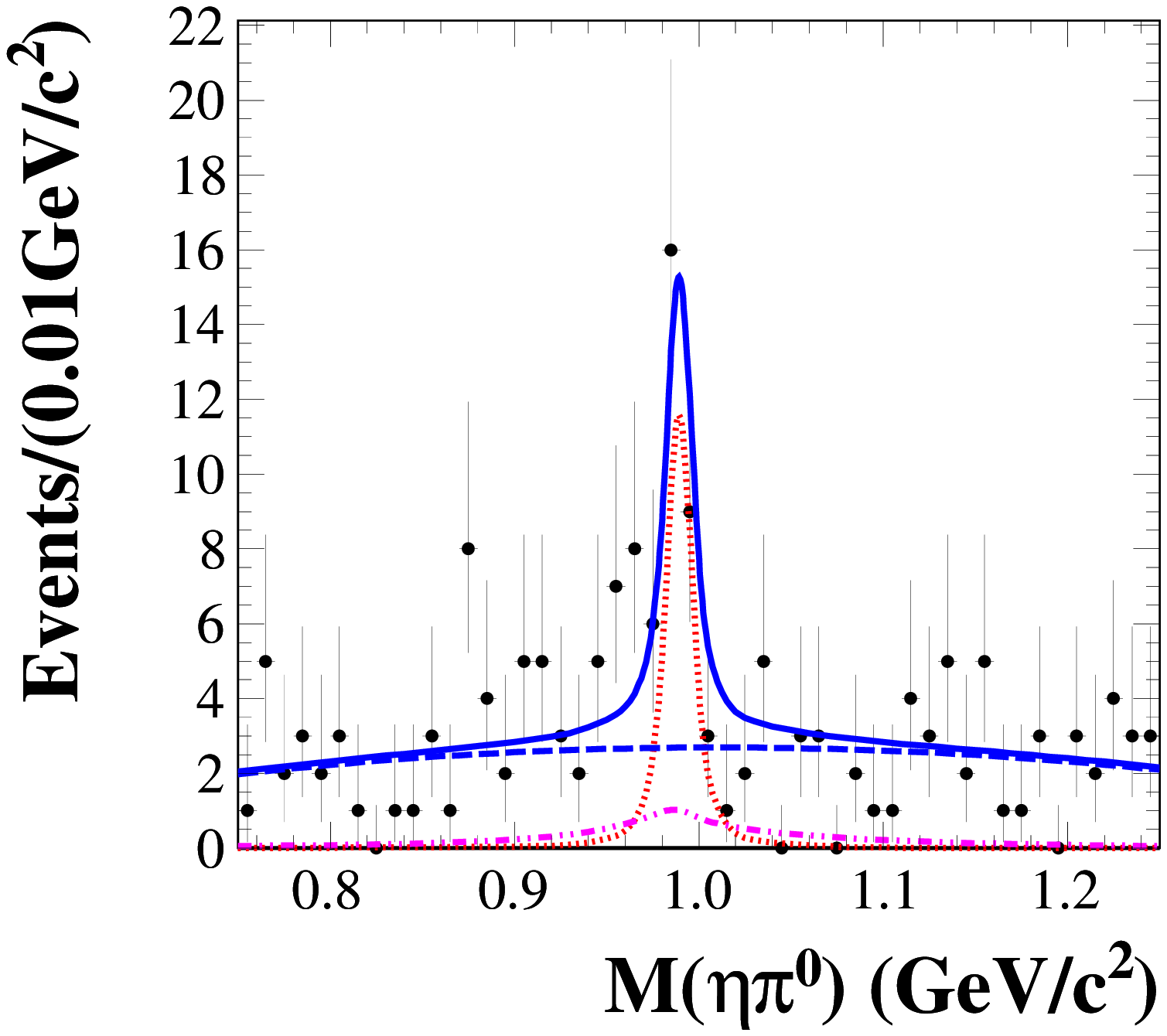}
  \put(-90,9){(a)}
  \includegraphics[width=4.5cm,height=4.5cm]{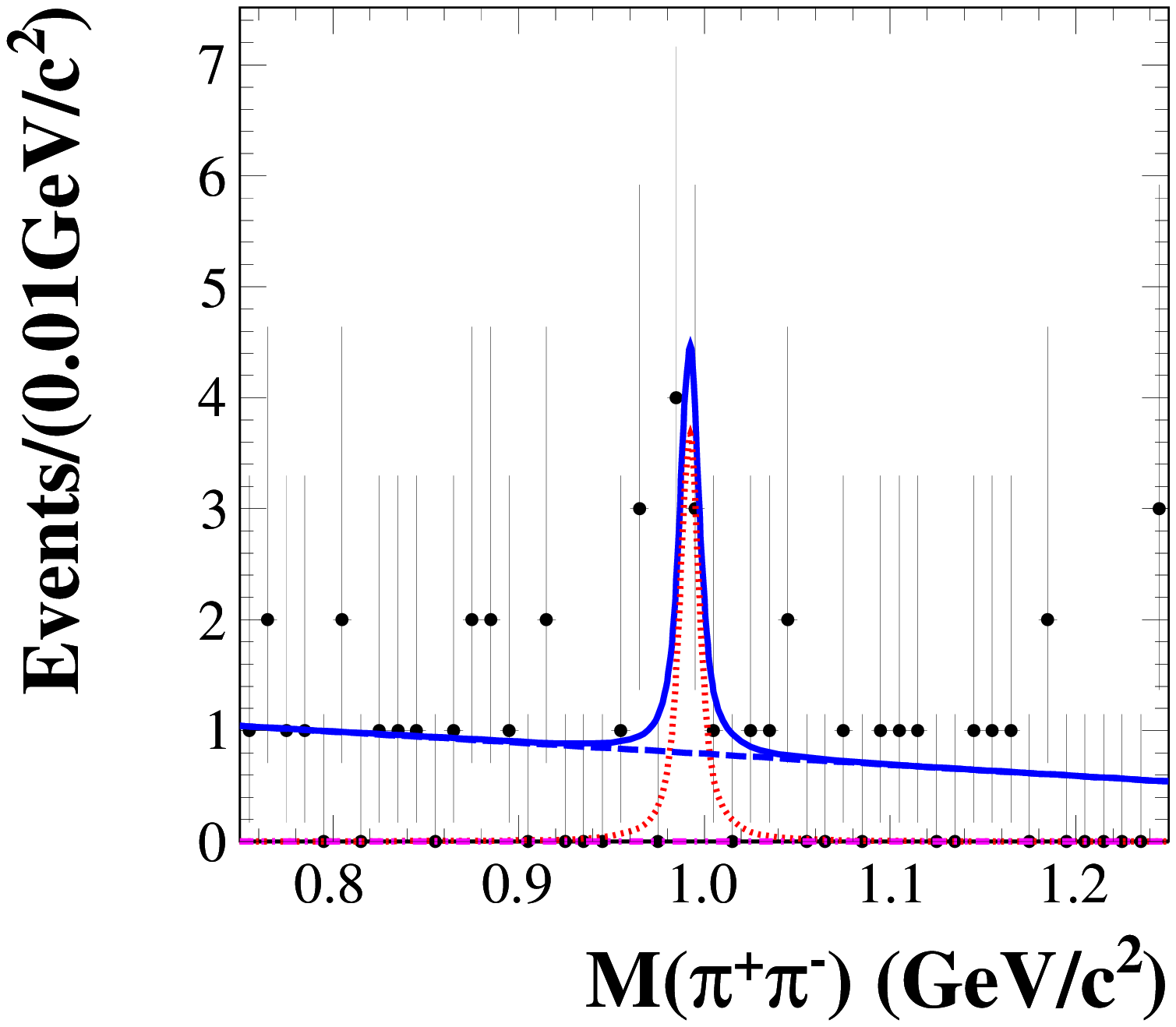}
  \put(-92,9){(b)}
\end{center}
\end{figure}
\\For $a^{0}_{0}(980)\to f_{0}(980)$ transition, the mass spectrum of $\pi^{+}\pi^{-}$ in the $\chi_{c1}$ mass window is fitted in a similar style, shown in FIG. (b).The significance of  $a^{0}_{0}(980)\to f_{0}(980)$ mixing signals is 1.9 $\sigma$.
The mixing branching ratio is determined to be $\br{\psi^{\prime}\to \gamma\chi_{c1}\to\gamma\pi^{0} a^{0}_{0}(980)\to\gamma\pi^{0}f_{0}(980)\to\gamma\pi^{0}\pi^{+}\pi^{-}}$($<6.0\times 10^{-7}$ at the $90\%$ C.L.).The mixing intensities $\xi_{fa}$ for $f_{0}(980)\to a^{0}_{0}(980)$ transition and $\xi_{af}$ for $a^{0}_{0}(980)\to f_{0}(980)$ transition are defined and calculated as:\\ \\$\xi_{fa} = \frac {\br{J/\psi\to\phi f_{0}(980)\to\phi a^{0}_{0}(980)\to\phi\eta\pi^{0}}} {\br{J/\psi\to\phi f_{0}(980)\to\phi\pi^{+}\pi^{-}}}$ ($<1.1\%$ at the $90\%$ C.L.),\\ \\
$\xi_{af} = \frac {\br{\psi^{\prime}\to\gamma\chi_{c1}\to\gamma\pi^{0} a^{0}_{0}(980)\to\gamma\pi^{0} f_{0}(980)\to\gamma\pi^{0}\pi^{+}\pi^{-}}} {\br{\psi^{\prime}\to\gamma\chi_{c1}\to\gamma\pi^{0} a^{0}_{0}(980)\to\gamma\pi^{0}\pi^{0}\eta}}$ ($<1.0\%$ at the $90\%$ C.L.).

\section{Summary}
A new facility for physics in the charm-$\tau$ region BEPCII/BESIII has become operational. With the world¡¯s largest samples of $J/\psi$ and $\psi^{\prime}$ collected at BESIII, the direct measurement of $a_{0}$ and $f_{0}$ mixing is performed for the first time.The measurements will be very helpful for pinning down the natures of $a^{0}_{0}(980)$ and $f_{0}(980)$.


%

}  


\end{document}